\documentstyle[epsfig,12pt]{article}
\def\csw{c_{SW}}
\def\k0{\kappa_0}
\def\kx{\kappa_\chi}
\def\kc{\kappa_{\rm crit}}
\def\g5{\gamma_5}
\def\bra{\langle}
\def\ket{\rangle}
\newcommand{\ba}{\begin{array}}
\newcommand{\ea}{\end{array}}
\newcommand{\bc}{\begin{center}}
\newcommand{\ec}{\end{center}}
\newcommand{\bd}{\begin{displaymath}}
\newcommand{\ed}{\end{displaymath}}
\newcommand{\be}{\begin{equation}}
\newcommand{\ee}{\end{equation}}
\newcommand{\bea}{\begin{eqnarray}}
\newcommand{\eea}{\end{eqnarray}}
\newcommand{\bi}{\begin{itemize}}
\newcommand{\ei}{\end{itemize}}
\newcommand{\btab}{\begin{tabbing}}
\newcommand{\etab}{\end{tabbing}}
\newcommand{\btbl}{\begin{table}}
\newcommand{\etbl}{\end{table}}

\def\1ad{\mbox{\normalsize $^1$}}
\def\2ad{\mbox{\normalsize $^2$}}
\begin {document}
\title{\bf
   Low-lying Eigenvalues of the improved Wilson-Dirac Operator in QCD
   \thanks{Talk presented at the 31st International Symposium
      Ahrenshoop on the Theory of Elementary Particles,
      Buckow, September 2-6, 1997
   }
}
\author{ \Large\bf Hubert Simma\1ad, Douglas Smith\2ad \\
               (UKQCD collaboration)
    \and
         \1ad \normalsize DESY-Zeuthen, Platanenallee 6, 
                        D-15738 Zeuthen, Germany
         \vspace*{2mm}
    \and
         \2ad \normalsize Dept. of Physics and Astronomy, University of Edinburgh,\\
              \normalsize Edinburgh EH9 3JZ, Scotland 
}

\date{}

\maketitle

\begin{abstract}
The spectral flow of the low-lying eigenvalues of the improved and unimproved 
Wilson-Dirac operator is studied on instanton-like configurations and on 
thermalized quenched configurations at various $\beta$-values and
lattice sizes. We also investigate the space-time localisation and 
chirality of the corresponding eigenvectors.
\end{abstract}

\section{Introduction}

The spectrum of the Dirac operator in QCD around zero is of particular 
physical interest, because in the continuum it is directly related to 
topological concepts and to the phenomenon of spontaneous breaking of 
chiral symmetry.

The topological charge of a smooth classical gauge field is related by the 
Atiyah-Singer index theorem to the difference between the number of right- 
and left-handed exact zero modes of the massless Dirac operator 
in that gauge background.
After performing the ensemble average of the path integral and in the limit of 
an infinite volume, the Banks-Casher formula relates the chiral condensate 
$\bra\bar\psi\psi\ket$ to the spectral density for zero eigenvalues of the 
Dirac operator \cite{BC}.
Moreover, in the framework of an effective chiral Lagrangian one finds
sum rules for moments of the spectral density in individual topological 
sectors, and the derivative of the spectral density at zero can be related 
to the decay constant of the pion \cite{LS}.

The situation is substantially more complicated in numerical simulations of 
QCD on finite discrete lattices. Besides the restriction to a finite volume 
and the difficulties of defining a topological charge, the use of the Wilson 
action explicitly breaks chiral symmetry.

In numerical simulations the low-lying eigenvalues of the Wilson-Dirac operator 
are also crucial from a practical point of view. In the
quenched approximation (almost-)zero modes are believed to be the origin 
of ``exceptional'' configurations, on which the numerical computation of 
light quark propagators is difficult and expensive, and which 
lead to large statistical errors of hadronic quantities at small
quark masses \cite{Eichten}.
In simulations with dynamical fermions the eigenvalues of the Wilson-Dirac
operator play a central r\^ole for the performance of the algorithms and 
their scaling behaviour when approaching the chiral limit \cite{KAlgo}.

In this work we investigate the appearance and space-time localisation 
of zero and low-lying modes of the hernitian Wilson-Dirac operator $\g5 M$.
In particular, we are interested in the effects of the Sheikholeslami-Wohlert
(SW) improvement term, the variation of the zero-mode distribution
with the gauge coupling, and the connection between the eigenmodes and
topological properties of the underlying gauge field.

This contribution is organised as follows: After the discussion of 
general properties of the spectra of $M$ and $\g5 M$ in the next section,
we investigate in sect.~3 the zero modes on simple, instanton-like gauge 
configurations, before and after heating, and with and without improvement.
In sect.~4, we present numerical data for thermalized quenched 
configurations on $16^4$ lattices, and discuss our results in sect.~5.

\section{Properties of the Eigenmodes of $\bf \g5 M$}

The (improved) Wilson-Dirac matrix is given by
\bea
    M  
    & = & 1 - \kappa ( H + \frac{i}{2} c_{SW} F_{\mu \nu} \sigma_{\mu \nu} )
    \ ,
\label{Dirac}
\eea
where $H$ is the usual hopping term,
and $\kappa$ is related to the bare quark mass $m_0$ by $\kappa=(8+2m_0)^{-1}$.
The boundary conditions for the fermion fields
are anti-periodic in time direction and periodic in the space directions.
The SW 
improvement term contains the usual clover combination $F_{\mu\nu}$ of the 
SU(3) gauge links.

Due to the $\g5$-hermiticity, $M^\dagger = \g5 M \g5$, the eigenvalues 
of $M$ come in complex conjugate pairs $\mu$ and $\mu^\ast$.
In addition, for $\csw=0$ the spectrum on even lattices is also reflection 
symmetric along the line ${\rm Re}\mu = 1$, i.e.
\bd
  M r = (1-\kappa \rho) r   
  ~~~ 
  \Leftrightarrow 
  ~~~
  M s = (1+\kappa \rho) s 
  \ ,
\ed
 with the eigenvectors related by the stagger transformation
$s(x) = (-1)^{\sum_\mu x_\mu}~r(x)$.

The elementary relation, 
$
(\mu_k^* - \mu_i) \cdot r_k^\dagger \gamma_5 r_i = 0
\ , \ 
$
between eigenvalues $\mu_i$ and corresponding right
eigenvectors $r_i$ of $M$,
suggests to distinguish two classes of eigenmodes:
\bi
\item ``bulk'' modes with $\mu \notin {\bf R}$ and vanishing chirality
      $(r,\gamma_5 r)$
\item real modes with $\mu \in {\bf R}$ and possibly non-zero chirality
\ei
All eigenvalues lie within the disk $\{\mu=1+\kappa\rho: \vert\rho\vert\le 8\}$.
On configurations with increasing $\beta$-values, the bulk modes depopulate
the vicinity of the real axis except for five regions corresponding to
the different corners of the Brillouin zone where the quark propagator
has poles.

One may expect that some of the real modes are related to topological 
properties of the gauge field by an approximate lattice remnant of 
the index theorem, but of course also topologically trivial gauge fields 
can have real modes (e.g. $U_4(x)= diag(1,-1,-1)$ on one time slice and all 
other $U_\mu(x)=\bf 1$). 

In the following we shall investigate eigenmodes of the 
{\em hermitian} operator
\be \label{q}
Q=\gamma_5 M/\left(1+8\kappa\right)\ .
\ee
The normalisation is such that the eigenvalues $\lambda(\kappa)$ of $Q$ 
satisfy $0\le \lambda^2 \le 1$, and we refer to the part of the
spectrum closest to zero as the ``low-lying'' eigenvalues.
They can be computed in an efficient and reliable way by using an 
accelerated conjugate gradient algorithm \cite{CG}, which also
provides approximate eigenvectors and is viable for large lattice sizes.

Recalling the relation $M^{-1} = Q^{-1} \gamma_5 /\left(1+8\kappa\right)$,
we note that all hadronic propagators can be constructed from $Q^{-1}$,
which has a simple spectral representation in terms of the orthonormal
eigenvectors of $Q$.

The eigenmodes (eigenvalues and corresponding eigenvectors) 
of $Q(\kappa)$ have a non-trivial $\kappa$-dependence.
The real modes, $\mu=1-\kappa\rho \in {\bf R}$, of $M$ are equivalent 
to zero-modes of $Q$ at $\kappa$-values $\k0=1/\rho$.
In general, the spectral flow $Q(\kappa)$ provides indirect information 
about the complex spectrum of $M$, and in particular the flow of the 
low-lying eigenvalues $\lambda(\kappa)$ probes the spectrum of $M$
close to the real axis. 
The five regions, where bulk modes of $M$ populate the vicinity of the 
real axis, correspond to minima of the low-lying eigenvalues of $Q(\kappa)$ 
at $\kappa$-values roughly of the order of
$0.125$, $0.25$, $\pm \infty$, $-0.25$, and $-0.125$.
In the special case of a flat gauge field each eigenvalue $\mu=1-\kappa\rho$ 
of $M$ corresponds to an extremal value in the spectral flow of 
$\tilde\lambda=\lambda\cdot(1+8\kappa)/\kappa$
at $\kappa=1/{\rm Re}\rho$.

The chiralities of the eigenvectors of $Q$
\be
   \chi_i \equiv (u_i, \gamma_5 u_i)
\ee
are related to the spectral flow of 
$\tilde\lambda_i(\kappa)$
by the identity  
\be
   \chi_i(\kappa) = \frac{d \tilde{\lambda}_i}{d (1/\kappa)}\ ,
\label{chi}
\ee
i.e. the zeroes of the chirality correspond to extrema in the spectral flow 
of $\g5 M/\kappa$. 
Eq.~(\ref{chi}) yields the first order $\kappa$-dependence of 
$\lambda(\kappa)$, and from the flow of a zero modes close to the position $\k0$,
where $\lambda$ vanishes, one can thus estimate $\k0$ as
\bd
   \kappa_0 \approx \kappa \cdot\left(1 + \frac{\chi}{\lambda-\chi}\right) \ ,
\ed
which is very helpful for an efficient iterative search for possible zero modes.

\section{Instanton-like Configurations}
We first investigate to what approximation the index theorem holds
for Wilson fermions on simple instanton-like configurations.
These are set up on even lattice sizes in the singular gauge as described 
in \cite{LatInst}, followed by a few cooling sweeps to reduce boundary 
mismatches. 
We consider configurations with up to two well-separated instantons
or anti-instantons in the same SU(2) subgroup and a net topological
charge $\nu=0,\pm1,\pm2$.

In the physical $\kappa$-region we find $n_+=\nu$ zero modes 
with chiralities of the same sign as $\nu$, and $n_-=0$ with
opposite sign chiralities. 
On multiple-instanton configurations their positions are split
depending on the size and distance of the instantons.
In the four unphysical regions one has $(n_+,n_-)=(0,4\nu)$, $(6\nu,0)$, 
$(0,4\nu)$, and $(\nu,0)$, respectively. In the following we consider only 
the physical region $0.1\le\kappa\le 0.2$. 

The spectral flow of the low-lying bulk modes on cold instanton-like 
configurations is characterised by a value of $\kx$ which is close to 
the free-case value of 1/8 and essentially independent of the instanton 
size 
$\rho$. On the other hand, the position $\k0$ and the chirality
$\chi_0(\k0)$ of the zero mode seem to be sensitive to discretisation 
effects and are strongly $\rho$-dependent. 
The following table shows the values for single instantions on a $16^4$ lattice:
\begin{center}
  \begin{tabular}{l|cc|cc|}
     & \multicolumn{2}{c|}{$\csw=0$} & \multicolumn{2}{c|}{$\csw=1$} \\ \hline
     $\rho$ & $\kappa_0$ & $\chi_0(\kappa_0)$ 
          & $\kappa_0$ & $\chi_0(\kappa_0)$ \\ \hline
     2 & .135 & .66 & .1258 & .986 \\
     4 & .127 & .90 & .1250 & .999 \\
     8 & .126 & .99 & .1250 & .999 \\
  \end{tabular}
\end{center}

We note the clear effect of the improvement term (using the tree-level value 
$\csw=1$): It significantly rises the chirality $\chi(\k0)$ of the zero mode 
and moves the position $\k0$ towards lower values and closer to $\kx$.

When the instanton-like 
configurations are heated by a moderate number of
update sweeps, the zero-mode remains, but the position $\k0$ changes and 
fluctuates for different heating trajectories. The heating also moves
the positions $\kx$, where the chiralities of the lowest bulk modes vanish, 
to values around $\kc(\beta)$, where $\beta$ is the value at which the 
heating was performed.
For the improved operator the variation of $\k0$ is much smaller,
and the central value is lower and significantly closer to $\kx$ than 
for $\csw=0$.

Also on the heated configurations the eigenvectors are centered on the 
instantons (except for $\kappa\ll\kx$), but the space-time extension of 
the eigenvector density at $\k0$ is not as claerly scaling with the size 
of the instanton as in the cold case.

\section{Quenched Configurations}
To study the spectrum of the Wilson-Dirac operator on quenched configurations, 
we used 10 thermalized $16^4$ configurations for each of the gauge couplings, 
$\beta=5.9$, $6.0$, and $6.1$. The spectral flows were computed for 
$\csw=0$ and for the non-perturbative value $\csw(\beta)$ of ref.~\cite{CSW}.

\begin{figure}
\begin{center}
\vspace*{-18mm}
\epsfig{figure=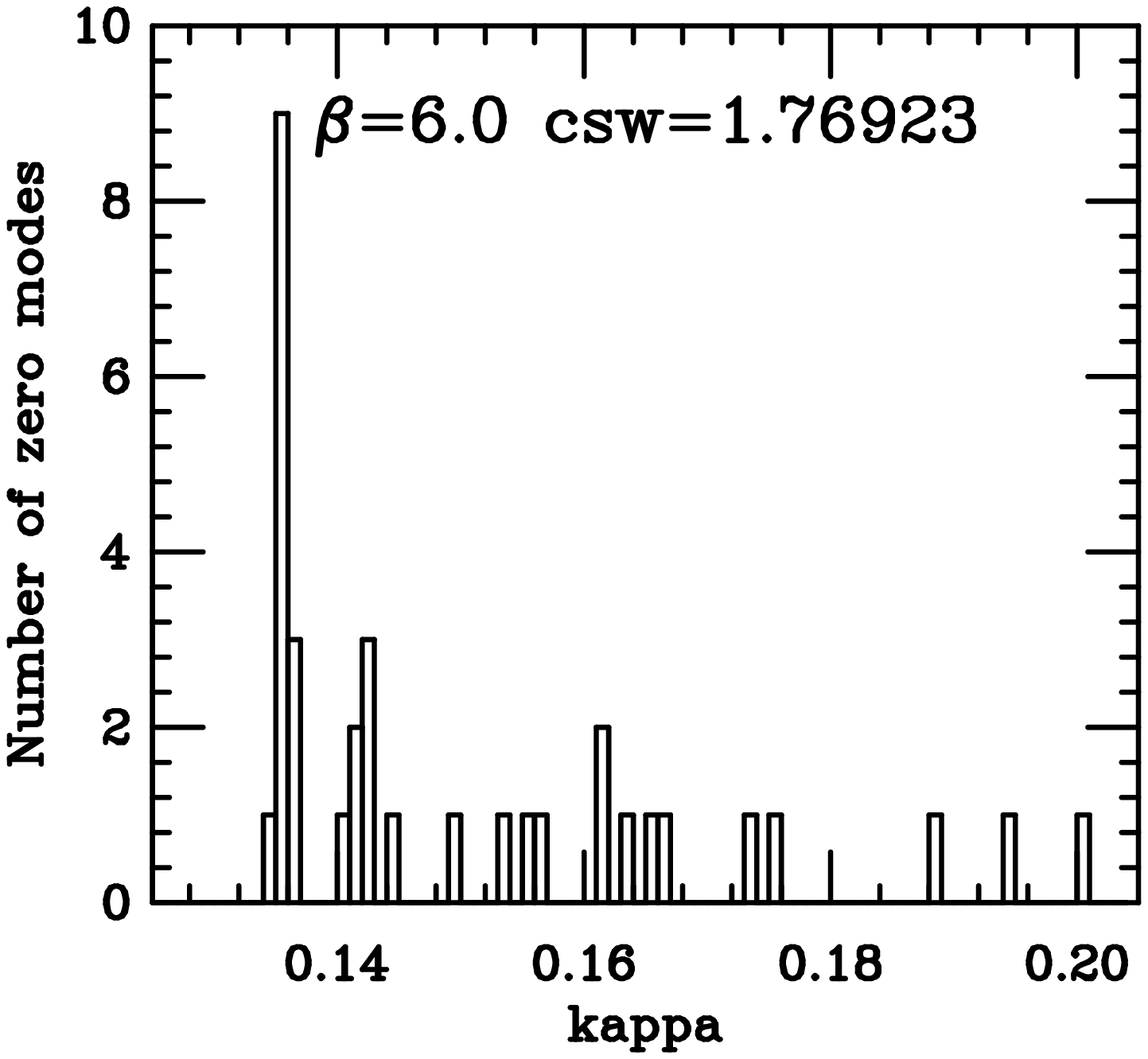,height=72mm}\hfill
\epsfig{figure=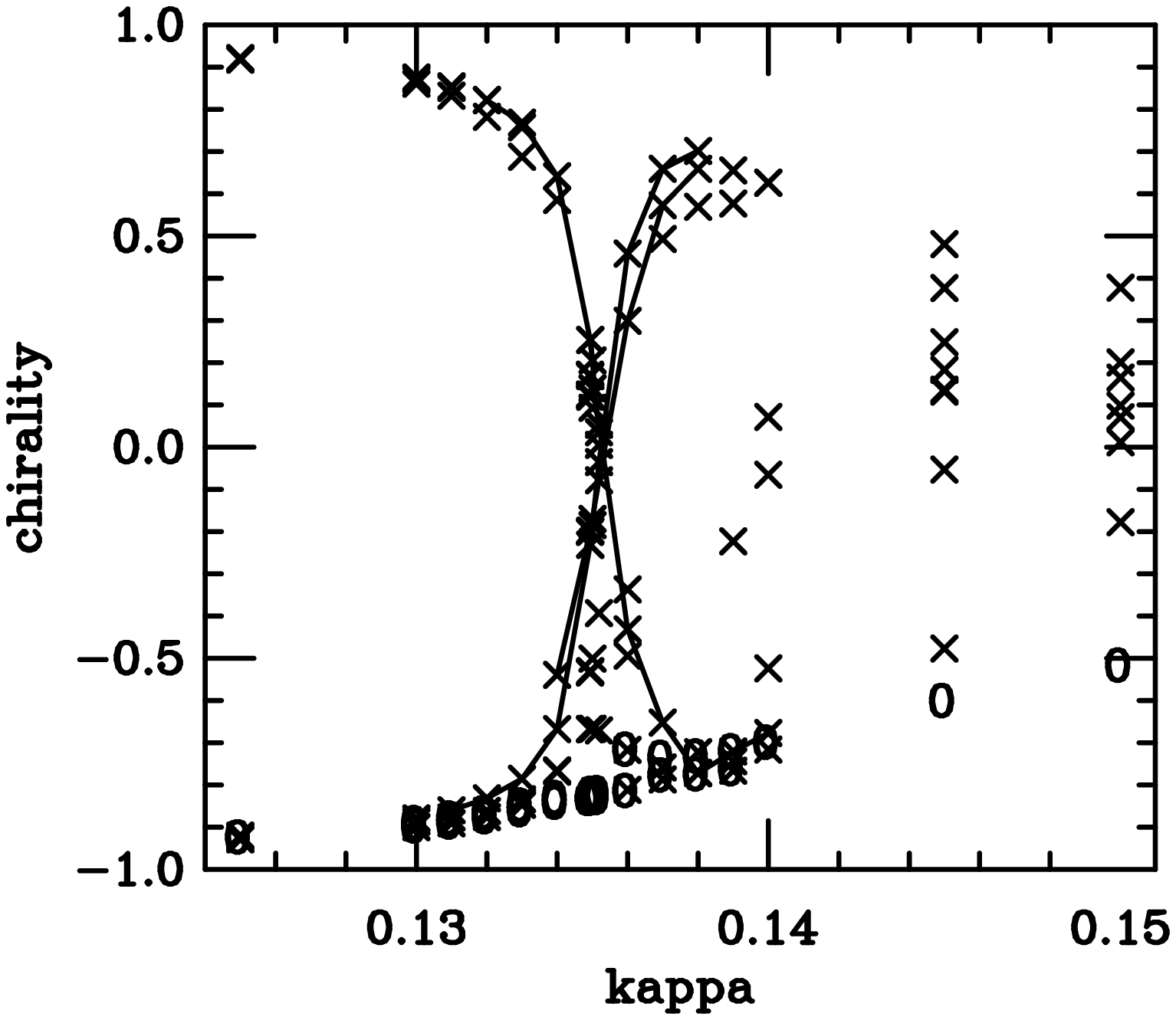,height=72mm}
\parbox{70mm}{
\vspace*{-0mm}
Figure~1: Distribution of the zero modes on 10 configurations at $\beta=6.0$.
}
\hfill
\parbox{70mm}{ 
\vspace*{-0mm}
Figure~2: Typical flow of the chirality of the lowest modes. 
}
\end{center}
\end{figure}

The distribution of the zero modes for the improved operator at $\beta=6.0$ 
is shown in Fig.~1. The peak of the distribution around $\kc$ is more pronounced
in the improved case than for $csw=0$.
The typical behaviour of the chiralities on one of these configurations is
illustrated in Fig.~2. The circles denote the lowest (zero) modes and the 
lines connect points from the same bulk modes to guide the eye. One 
notices that the chiralities of the low-lying bulk modes vanish around 
the same value $\kx\approx\kc(\beta)$.

The space-time localisation of the low-lying eigenmodes for the unimproved 
case, is known to change around $\kc$ from a typically exponential decay of 
the eigenvector density to an even stronger, almost point-like, localisation
\cite{lat96}.
For the improved operator we find in the region $\kappa\ll\kx$ a somewhat
stronger localisation of the lowest-lying modes than for $\csw=0$, 
but this difference tends to disappear towards $\kx$. Above $\kx$ 
a sharp drop of the ``participation ratio'' indicates again a very strong 
(almost point-like) localisation similar to the unimproved case. 
The $\kappa$-dependence of the localisation strength is the same for the 
lowest-lying bulk modes and for the flow of zero modes. Thus, zero modes 
appearing at very low $\k0\le\kx$ are in general weaker localised than those 
at higher $\k0$.

To investigate the localisation of the zero modes in relation to topological
properties of the underlying gauge field, we approximate the charge density
of the cooled gauge configurations by a superposition of single-instanton
charge distributions \cite{CoolCharge}. 
We find that the peaks closest to the maximum of the eigenvector
density are in average significantly higher (i.e. correspond
to smaller instantons) than the average value of the configuration. 
There also seems to exist a weak correlation between the chirality of the 
eigenmode and the sign of the topological charge of the closest peak.

\section{Discussion}

On simple instanton-like configurations we confirmed that the improved and 
unimproved Wilson-Dirac operator approximately satisfy an index theorem 
similar to the case of staggered fermions or QED$_2$ \cite{LatIndex}.
The presence of the zero modes is rather stable under the effect of 
roughening the gauge fields by heating.

Inclusion of the SW improvement term significantly
increases the chirality of the zero modes and moves $\k0$ closer to $\kc$.
On thermalized quenched configurations the improvement term renders the
zero mode distribution stronger peaked around $\kc$. This is in
accordance with the picture that in the continuum limit
the $\kappa$-region with vanishing mass gap (i.e. with zero modes of $Q$)
shrinks to a point \cite{SCRI}.

The number of zero modes found on the $16^4$ lattices clearly decreases 
with increasing $\beta$-values. This, and the much lower abundance
of zero modes on $8^4$ lattices, indicates that the number of zero
modes grows with the physical volume.

The eigenvectors of the improved operator are somewhat stronger localised 
than in the unimproved case, and their positions tend to lie close to
peaks of the topological charge density of the (cooled) gauge field.
The width of these peaks is typically smaller than the average for the
configuration.

On the four exceptional configurations encountered by UKQCD at $\beta=6.0$ on
$16^348$ and $32^364$ lattices we verified the presence of a zero mode at
a near $\kappa$-value $\k0<\kc$. In all cases the eigenvector is localised
close to or on a narrow peak in the topological charge density, and the 
chirality has the same sign as the charge of the peak, but 
is not correlated with the overall topological charge of the configuration.

The spectral flow of $Q(\kappa)$ can also provide instructive information
about the complex spectrum of $M$. For instance, the position $\kx$ of 
the chirality zeroes of the lowest bulk modes approximately coincides
with $\kc$ (defined as an ensemble quantity, e.g. from the pion mass),
and may be an interesting quantity to characterise the ``critical''
$\kappa$-value for individual gauge configurations. 

Further details and numerical data shall be presented elsewhere
\cite{EVpaper}.

\vspace*{5mm}
Thanks are due to the organisers for an interesting and pleasant
Symposium. We thank M.~B\"aker, K.~Jansen, M.~L\"uscher and S.~Pickles 
for helpful discussions, and R.~Petronzio for generous computing time 
on the APE at Tor Vergata. Part of the numerical computations
was also performed at DESY-Zeuthen, at the University of Edinburgh, and 
at the APE Lab, and we are grateful for their hospitality and support.


\begin{thebibliography}{77}

\bibitem{BC}
 T.~Banks and A.~Casher, {\it Nucl. Phys.} {\bf B169} (1980) 103;\\
 E.~Marinari, G.~Parisi, and C.~Rebbi, {\it Phys. Rev. Lett.} {\bf 37} (1981) 1795

\bibitem{LS}
 H.~Leutwyler, and A.~Smilga, {\it Phys. Rev.} {\bf D46} (1992) 5607;\\
 A.~Smilga, hep-th/9503049

\bibitem{Eichten}
 W.~Bardeen, et al., hep-lat/9710084 and hep-lat/9705002

\bibitem{KAlgo}
 K.~Jansen, {\it Nucl. Phys.} {\bf B} (Proc. Suppl.) {\bf 53} (1997) 262

\bibitem{CG}
 T.~Kalkreuter and H.~Simma, {\it Comp. Phys. Comm.} {\bf 93} (1996) 33.

\bibitem{LatInst}
 D.J.R.~Pugh and M.~Teper, {\it Phys. Lett.} {\bf B212} (1989) 326; \\
 M.~Laursen, J.~Smit, and J.~Vink, {\it Nucl. Phys.} {\bf B343} (1990) 522

\bibitem{CSW}
 M.~L\"uscher, et al., {\it Nucl. Phys.} {\bf B491} (1997) 323

\bibitem{lat96}
 K.~Jansen, et al. 
 {\it Nucl. Phys.} {\bf B} (Proc. Suppl.) {\bf 53} (1997) 262

\bibitem{CoolCharge}
 D.~Smith and M.~Teper (UKQCD Collaboration), hep-lat/9801008

\bibitem{LatIndex}
 I.~Barbour, and M.~Teper, {\it Phys. Lett.} {\bf B175} (1986) 445;\\
 S.~Itoh, Y.~Iwasaki, and T.~Yoshi\'e, {\it Phys. Rev.} {\bf D36} (1987) 527;\\
 J.~Smit, and J.~Vink, {\it Nucl. Phys.} {\bf B286} (1987) 485;\\
 C.~Gattringer, I.~Hipp, and C.~Lang, 
 {\it Nucl. Phys.} {\bf B508} (1997) 329, and hep-lat/9712015 

\bibitem{SCRI}
 R.G.~Edwards, U.M.~Heller, R.~Narayanan, R.L.~Singleton, hep-lat/9711029

\bibitem{EVpaper}
 H.~Simma, in preparation

\end{thebibliography}
\end{document}